\documentclass[12pt]{article}
\usepackage{graphicx}
\usepackage{lineno}

\newcommand\pubnumber{SNSN-323-63}
\newcommand\pubdate{\today}

\def\institute{CERN\\
CH-1211 Geneva 23, Switzerland}

\def\Title#1{\begin{center} {\Large #1 } \end{center}}
\def\Author#1{\begin{center}{ \sc #1} \end{center}}
\def\Address#1{\begin{center}{ \it #1} \end{center}}

\newcommand\pubblock{\rightline{\begin{tabular}{l} \pubnumber\\
         \pubdate  \end{tabular}}}
\newenvironment{Abstract}{\begin{quotation}  }{\end{quotation}}
\newenvironment{Presented}{\begin{quotation} \begin{center} 
             PRESENTED AT\end{center}\bigskip 
      \begin{center}\begin{large}}{\end{large}\end{center} \end{quotation}}

\def\beq{\begin{equation}}
\def\eeq#1{\label{#1}\end{equation}}
\def\eeqn{\end{equation}}

\def\beqa{\begin{eqnarray}}
\def\eeqa#1{\label{#1}\end{eqnarray}}
\def\eeqan{\end{eqnarray}}

\let\bar=\overbar

\def\Dslash{\not{\hbox{\kern-4pt $D$}}}
\def\dslash{\not{\hbox{\kern-2pt $\del$}}}

\def\msb{{\bar{\ssstyle M \kern -1pt S}}}

\begin{document}
\begin{titlepage}
\pubblock

\vfill
\Title{Measurements of inclusive and differential fiducial cross-sections of $t\bar{t}\gamma$ production in leptonic final states at $\sqrt{s}$~=~13~TeV}
\vfill
\Author{Mar\'ia Moreno Ll\'acer, on behalf of the ATLAS Collaboration}
\Address{\institute}
\vfill
\begin{Abstract}
The latest results of the ATLAS experiment for the production of a top-quark pair in association with a photon using proton-proton collision data from the LHC at a centre-of-mass energy of 13~TeV are summarised here. This includes inclusive and differential cross-sections measurements performed in single-lepton and dilepton final states in a fiducial volume corresponding to the experimental acceptance.
\end{Abstract}
\vfill
\begin{Presented}
$11^\mathrm{th}$ International Workshop on Top Quark Physics\\
Bad Neuenahr, Germany, September 16--21, 2018
\end{Presented}
\vfill
\small 
\copyright 2019 CERN for the benefit of the ATLAS Collaboration.\\
Reproduction of this article or parts of it is allowed as specified in the CC-BY-4.0 license.
\end{titlepage}
\def\thefootnote{\fnsymbol{footnote}}
\setcounter{footnote}{0}

\section{Introduction}

The production of a top-quark pair in association with a photon is a direct probe of the electromagnetic coupling of the top quark. Anomalous top-quark couplings could manifest as discrepancies with respect to the Standard Model (SM) predictions in the shapes of various kinematic distributions or in absolute cross-section measurements.
The photon can originate not only from a top quark, but also from its charged decay products, including charged fermions (quark or lepton) from the $W$~boson decay. In addition, it can be radiated from a charged incoming parton. In the analysis discussed here~\cite{ATLAS_ttgamma13TeV}, no attempt is made to separate these different sources of photons, but an event selection is applied to suppress those radiated from top-quark decay products. 
Fiducial cross-sections are measured in the single-lepton and dilepton channels using proton-proton collision data corresponding to an integrated luminosity of 36.1~fb$^{-1}$ and collected by the ATLAS detector~\cite{ATLAS} at the LHC in 2015-2016 at a centre-of-mass energy of 13~TeV~\cite{ATLAS_ttgamma13TeV}. In each channel, the fiducial cross-section is measured with a likelihood fit to the output of a neural network (NN) trained to differentiate between signal and background events. In both channels, normalised differential cross-sections are measured in the same fiducial region as a function of the photon $p_\mathrm{T}$, the photon $|\eta|$ and the angular distance $\Delta$$R$ between the photon and its closest lepton\footnote{ATLAS uses a right-handed coordinate system with its origin at the nominal interaction point (IP) in the center of the detector and the $z$-axis along the beam pipe. The $x$-axis points from the IP to the center of the LHC ring, and the $y$-axis points upward. Cylindrical coordinates $(r,\phi)$ are used in the transverse plane, $\phi$ being the azimuthal angle around the $z$-axis. The pseudorapidity is defined in terms of the polar angle $\theta$ as $\eta=-\ln\tan(\theta/2)$. The angular distance $\Delta$$R$ between two objects is defined as  $\Delta R \equiv \sqrt{(\Delta \eta)^2 + (\Delta\phi)^2}$.}. In the dilepton channel, these are also measured as a function of the absolute pseudorapidity distance $\Delta\eta$ and the azimuthal angle $\Delta\phi$ between the two leptons. The measured cross-sections are compared to predictions from leading order generators at particle level (stable particles before detector simulation). The predictions for the inclusive cross-sections are corrected by next-to-leading order (NLO) k-factors in the strong interaction~\cite{Kfactors_ttgamma13TeV}, calculated at parton level.

\section{Event selection}
Candidate events were collected using single-lepton triggers. Electrons and muons are both required to be isolated based on calorimeter and track information and to have $p_\mathrm{T}>~27.5$~GeV, except for electrons in 2015 data for which this cut is lowered to $p_\mathrm{T}>~25$~GeV.
The selected events must have at least four (two) jets in the single-lepton (dilepton) channel, at least one of which should be tagged as a $b$-jet with a nominal efficiency of 77$\%$, and exactly one photon. A $Z$-boson veto is applied in the single-electron channel by excluding events with invariant mass of the system of the electron and the photon within 5~GeV of the $Z$-boson mass. In the dilepton channel, when the two leptons have the same flavour, events are excluded if the dilepton invariant mass or the invariant mass of the system of the two leptons and the photon is between 85 and 95 GeV, and the missing transverse momentum $E_{\mathrm{T}}^{\mathrm{miss}}$ is required to be larger than 30~GeV. The dilepton invariant mass is required to be higher than 15 GeV to suppress low-mass Drell-Yan events. Finally, to suppress photons radiated from leptons, the $\Delta$$R$ between the selected photon and any lepton must be greater than 1.0.

The fiducial phase space is defined at particle level such that it mimics the above event selections. In general, the selections are the same. However, cuts on invariant mass variables and $E_{\mathrm{T}}^{\mathrm{miss}}$ are removed.

A total number of 11662 and 902 candidate events are selected for the single-lepton and dilepton channels respectively, with an expected number of 6490$\pm$420 and 720$\pm$34 signal events, where the corresponding NLO k-factors are applied and both the simulation statistical uncertainty and systematic uncertainties are included.


\section{Background estimation}

There are four types of backgrounds to the selected $t\bar{t}$ candidates: one background is prompt-photon radiation and the other three are events with a fake (i.e. misidentified) object. The latter are contributions from events with the selected photon being:
\begin{itemize}
\item a misidentified jet or a non-prompt photon from hadron decays (\textit{hadronic-fake photon background}) [relevant in single-lepton channel]: these are mainly coming from $t\bar{t}$~events and estimated using a data-driven method, called the ABCD method, in which scale factors to correct Monte Carlo (MC) prediction to data are derived in the single-lepton channel using the ABCD method.  This set of scale factors is derived in the single-lepton channel and applied to both the single-lepton and dilepton channels to calibrate the simulation to match data.
\item a misidentified electron (\textit{electron-fake photon background}): this is mainly coming from $t\bar{t}$~dileptonic events and also estimated using a data-driven method in which scale factors derived from a data-driven tag-and-probe method are used to correct the number of fake photons predicted by the MC samples.\\
\end{itemize}
or with the selected lepton being:
\begin{itemize}
\item a misidentified jet or non-prompt lepton from heavy-flavour decays (\textit{fake-lepton background}) [relevant in single-lepton channel]: these are mainly coming from QCD multi-jet plus photon processes and estimated using the data-driven Matrix Method approach.
\end{itemize}
The contribution from events with prompt photon radiation not coming from the top quark or its decay products, referred to as \textit{prompt-photon background}, includes $W\gamma$, $Z\gamma$ and associated production of a photon in single top, diboson, and $t\bar{t}V$ production, the first two being the dominant ones in the single-lepton and dilepton channel respectively. The normalisation of the $W\gamma$ background is treated as a free parameter of the likelihood fit in the single-lepton channel, since this background is well separated from the $t\bar{t}\gamma$ signal and the uncertainty of its theoretical prediction is large. The shape of $W\gamma$ is taken from simulation and checked in the validation region to ensure good modelling. The normalisation and shape of the $Z\gamma$ background in the dilepton channel as well as other prompt backgrounds in both channels are predicted by simulation.

\begin{figure}[h!]
\centering
\includegraphics[height=6.5cm]{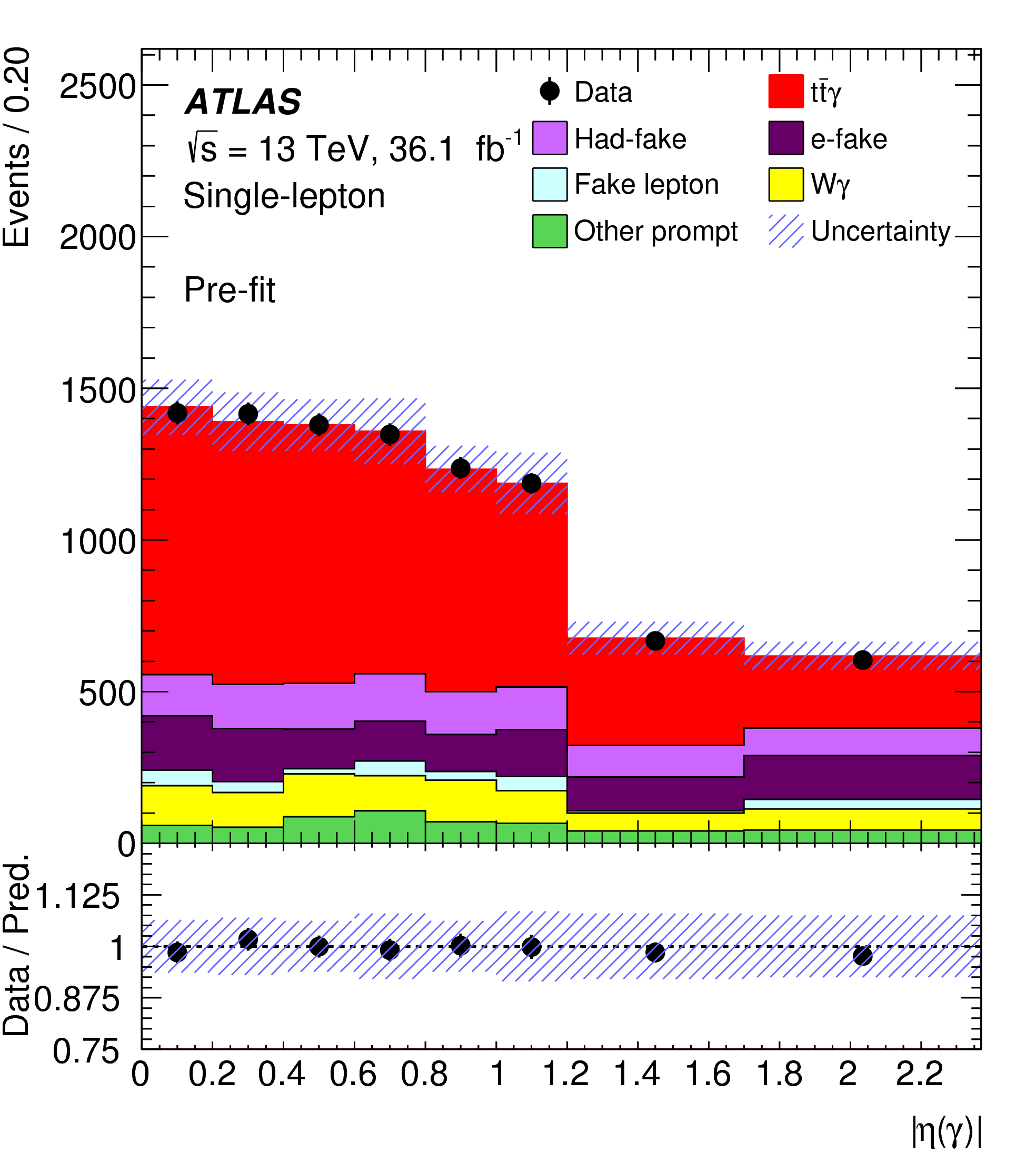}
\includegraphics[height=6.5cm]{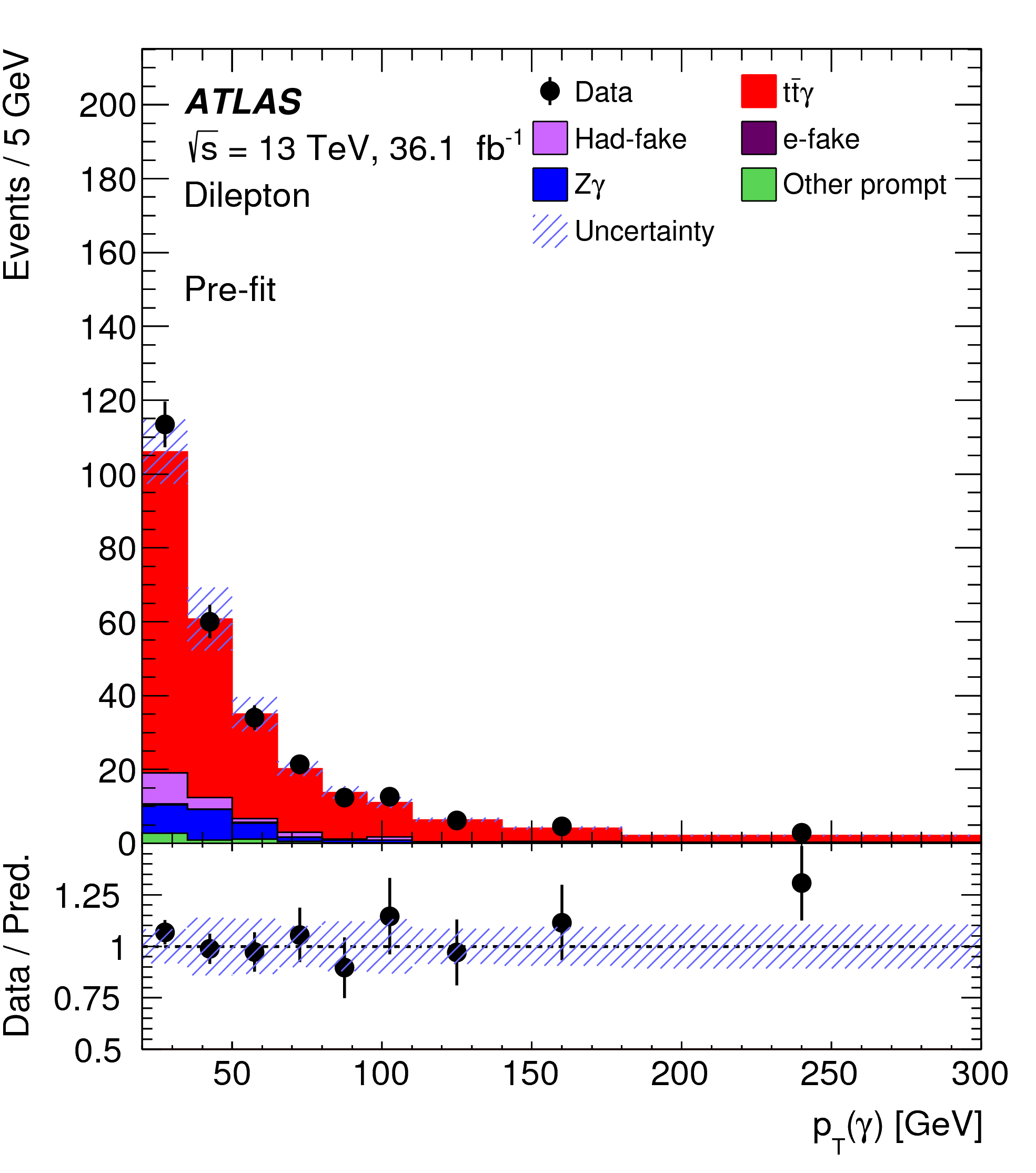}
\caption{Distributions of the photon $|\eta|$ ($p_\mathrm{T}$) in the single-lepton (dilepton) channel after event selection and before likelihood fit~\cite{ATLAS_ttgamma13TeV}. All data-driven corrections and systematic uncertainties are included.}
\label{fig:KineDist_prefit}
\end{figure}

Figure~\ref{fig:KineDist_prefit} shows some of the distributions of the observables before likelihood fit that are used for the measurement of the normalised differential cross-sections in the single-lepton and dilepton channels, respectively. All data-driven corrections and systematic uncertainties are included, and the signals are scaled by the NLO k-factors.

\section{Analysis strategy and results}
\subsection{Fiducial cross-section}
\begin{figure}[htb]
\centering
\includegraphics[height=6.5cm]{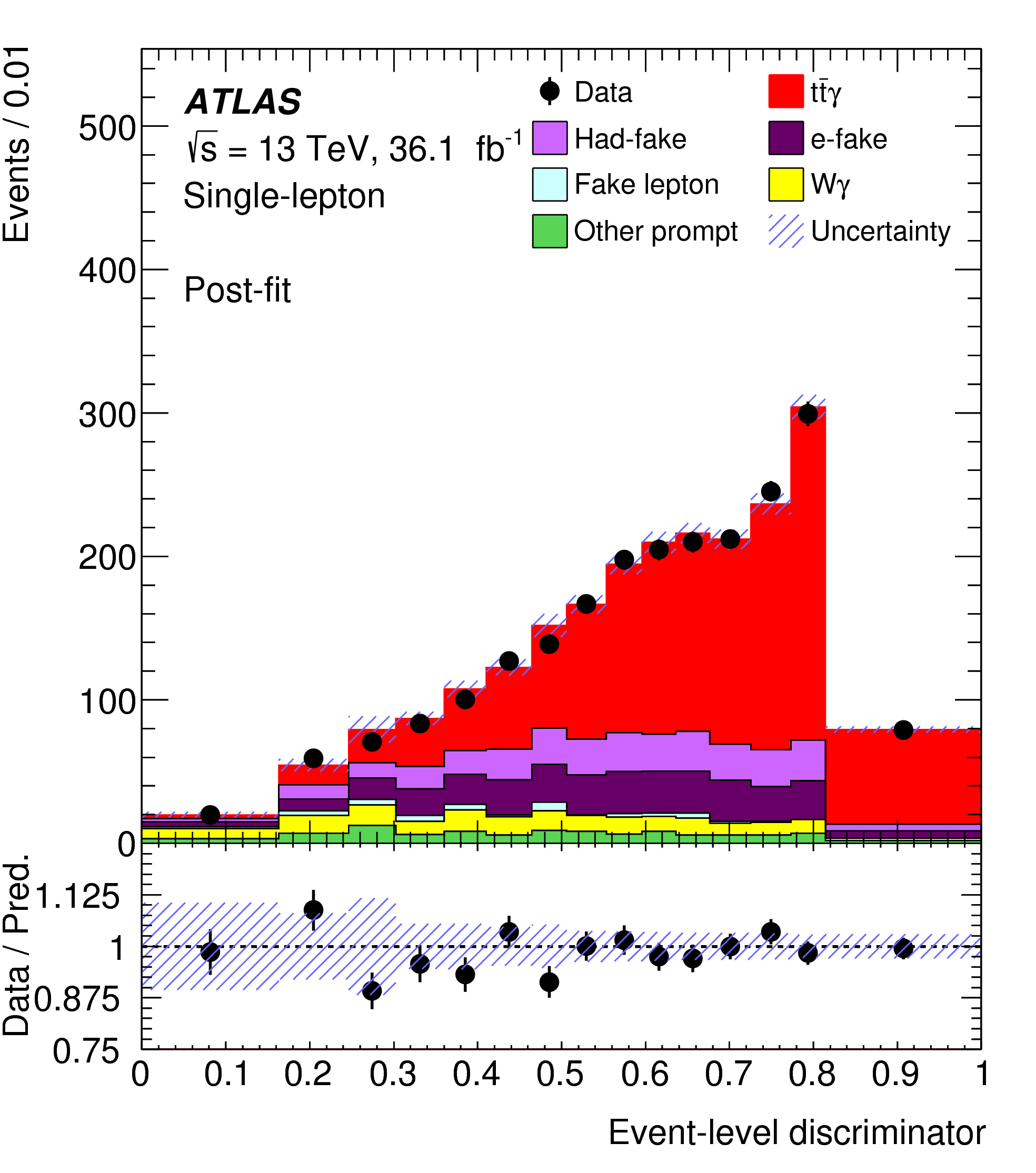}
\includegraphics[height=6.5cm]{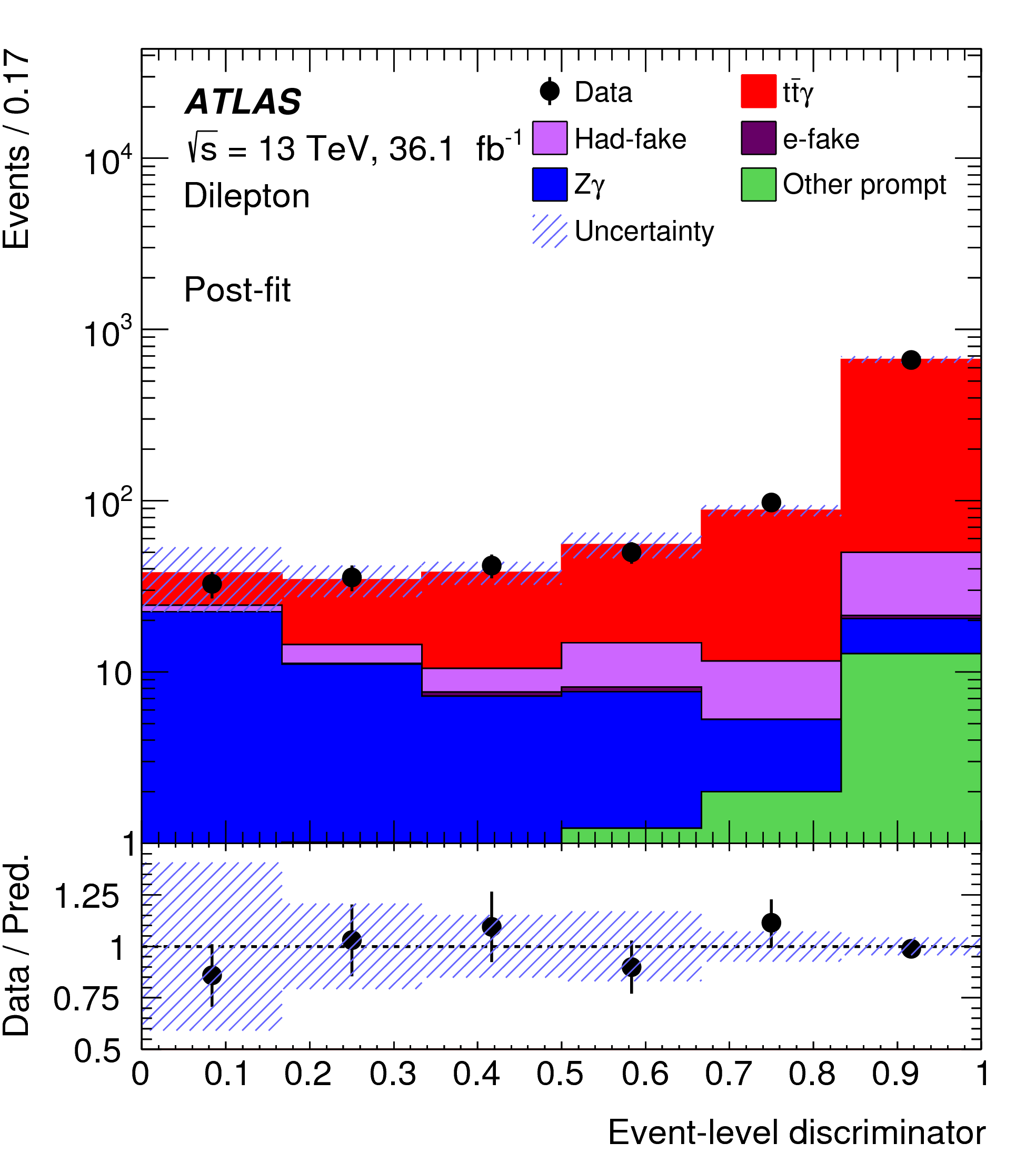}
\caption{The post-fit ELD distributions for the single-lepton and dilepton channels~\cite{ATLAS_ttgamma13TeV}. All the systematic uncertainties are included.}
\label{fig:ELD_postfit}
\end{figure}

Object-level and event-level neural networks are used to improve the sensitivity of the measurement in the single-lepton channel, while only an event-level NN is used in the dilepton channel.
The object-level NN, called the \textit{Prompt Photon Tagger}, is used to distinguish between prompt photons and hadronic fake photons and it is trained on independent prompt photon and dijet samples, using variables which characterize the photon candidate shower shape in the calorimeters. This NN serves as an input into the event-level NN, called the \textit{Event-Level Discriminator} (ELD) in the single-lepton channel, and it is one the most powerful variables together with $b$-tagging information. For the dilepton ELD, the $b$-tagged jet and the invariant mass between leptons prove to be very powerful.
The fiducial cross-sections for each channel are extracted by performing a maximum likelihood fit on the ELD and the measured values are found to be 521~$\pm$~9(stat.)~$\pm$~41(sys.)~fb and 69~$\pm$~3(stat.)~$\pm$~4(sys.)~fb for the single-lepton and dilepton channels, respectively. The corresponding post-fit ELD distributions are shown in Figure~\ref{fig:ELD_postfit}.


\subsection{Normalised differential cross-sections}
\begin{figure}[htb]
\centering
\includegraphics[height=6.5cm]{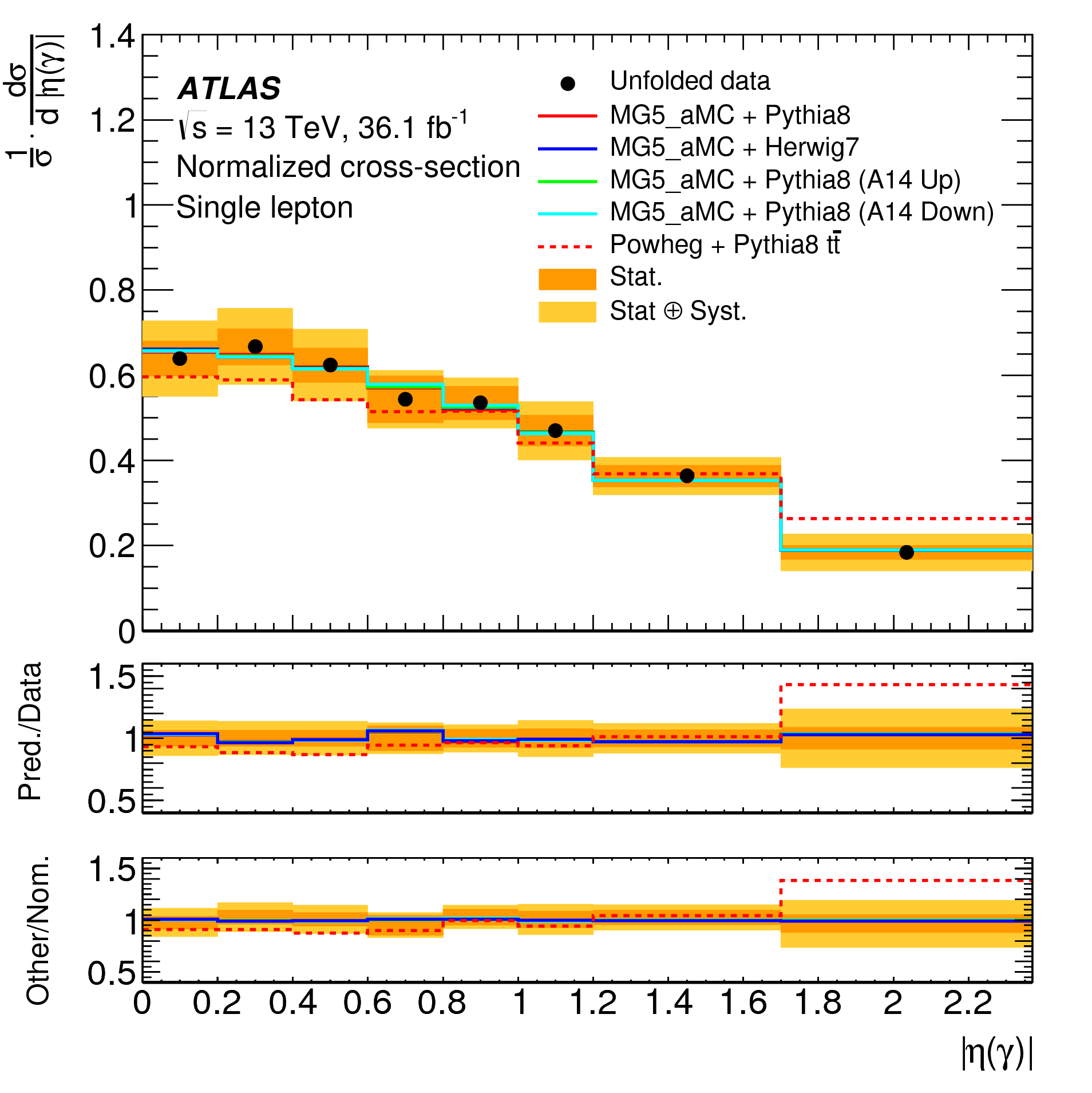}
\includegraphics[height=6.5cm]{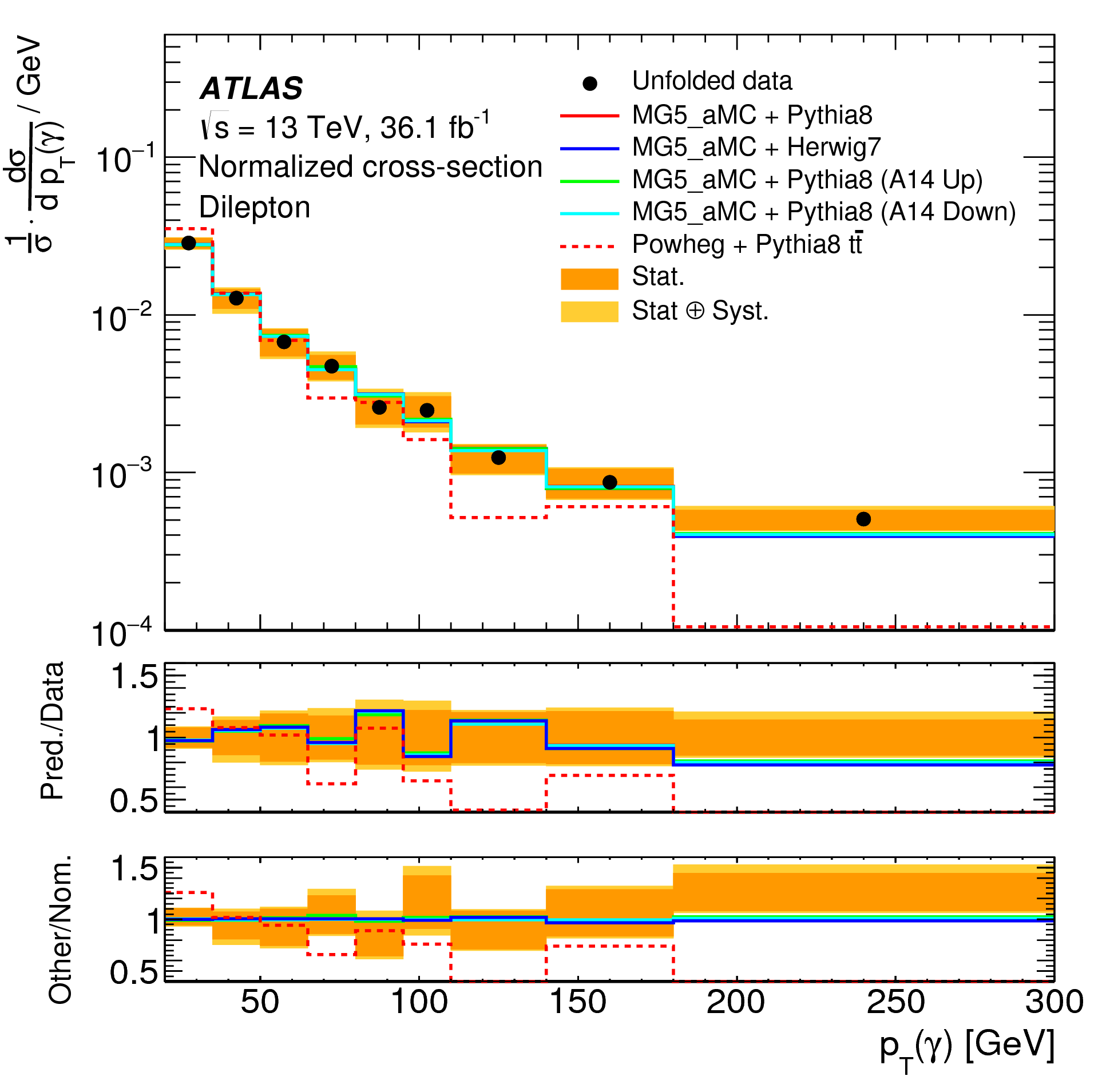}
\caption{The normalised differential cross-sections as a function of the photon $p_\mathrm{T}$ ($|\eta|$) in the single-lepton (dilepton) channel~\cite{ATLAS_ttgamma13TeV}. The unfolded distributions are compared to the predictions of MadGraph5$\_$aMC@NLO+Pythia8 together with the up and down variations of the Pythia8 A14 tune parameters, MadGraph5$\_$aMC@NLO+Herwig7, and Powheg+Pythia8 $t\bar{t}$ where photon radiation is modelled in the parton shower. The top ratio-panel shows the ratios of all the predictions over data. The bottom ratio-panel shows the ratios of the alternative predictions and data over the nominal prediction. Overflows are included in the last bin.}
\label{fig:KineDist_diffXS}
\end{figure}

The normalised differential cross-sections as a function of photon absolute pseudorapidity (transverse momentum) in the single-lepton (dilepton) channel are shown in Figure~\ref{fig:KineDist_diffXS}. An unfolding procedure is applied to the observed detector-level distributions in Figure~\ref{fig:KineDist_prefit}, after subtracting the background contributions. All measurements are found to be in agreement with the theoretical predictions.

\end{document}